# An orbitally derived single-atom magnetic memory

Brian Kiraly[1], Alexander N. Rudenko[2,1,3], Werner M.J. van Weerdenburg[1], Daniel Wegner[1], Mikhail I. Katsnelson[1], Alexander A. Khajetoorians[1*]

1. Institute for Molecules and Materials, Radboud University, Nijmegen, The Netherlands
2. School of Physics and Technology, Wuhan University, Wuhan 430072, China
3. Theoretical Physics and Applied Mathematics Department, Ural Federal University, 620002 Ekaterinburg, Russia

* Correspondence to: a.khajetoorians@science.ru.nl

## Abstract

A single magnetic atom on a surface epitomizes the scaling limit for magnetic information storage. Indeed, recent work has shown that individual atomic spins can exhibit magnetic remanence and be read out with spin-based methods – demonstrating the fundamental requirements for magnetic memory. However, atomic spin memory has been only realized on thin insulating surfaces to date, removing potential tunability via electronic gating or distance-dependent exchange-driven magnetic coupling. Here, we show a novel mechanism for single-atom magnetic information storage based on bistability in the orbital population, or so-called valency, of an individual Co atom on semiconducting black phosphorus (BP). Distance-dependent screening from the BP surface stabilizes the two distinct valencies and enables us to electronically manipulate the relative orbital population, total magnetic moment and spatial charge density of an individual magnetic atom without a spin-dependent readout mechanism. Furthermore, we show that the strongly anisotropic wavefunction can be used to locally tailor the switching dynamics between the two valencies. This orbital memory derives stability from the energetic barrier to atomic relaxation and demonstrates the potential for high-temperature single-atom information storage.

**Introduction**

Single-atom memory represents the ultimate limit in high-density storage[1-3] and a route toward quantum coherent manipulation[4,5]. Of particular interest are single magnetic atoms on surfaces, which can represent a bit employing the bistability of the magnetic moment, as they offer tunable interatomic coupling and bottom-up design[6]. While atomic spins can have long lifetimes[2,3,7], the key challenge has been to decrease fluctuations induced by spin-sensitive readout or scattering mechanisms[8,9] utilizing robust magnetic anisotropy[2]. The strategy toward single-atom magnetic memory has largely been to utilize certain combinations of atoms and surfaces in which the moment-bearing orbitals responsible for magnetism weakly hybridize with the environment[3,7]. Ultimately, this limits the selective coupling between neighboring atoms[10] and electronic access to the spin[11]. Single dopants in semiconductors[12] offer an attractive route toward atomic-scale memory[4] and processing[13,14], with the advantage of gating while still weakly hybridizing with the localized spin[15]. Scanning tunneling microscopy (STM) of individual non-magnetic dopants[16-18] and dangling bonds[19] revealed bistability of ionization states and has been used to study the interplay between local screening and confinement[20,21]. However, the magnetic properties of individual magnetic impurities in semiconductors[15] have been poorly explored compared to their counterparts on metals and thin-film insulators[22,23].

Black phosphorus (BP) is a particularly intriguing semiconductor and tunable Dirac material[24,25], with a strongly anisotropic band structure and thickness-dependent band gap[26]. To date, the magnetic properties of individual dopants in black phosphorus have not been experimentally investigated, in contrast to other Dirac materials[27,28]. Here, we demonstrate a single-atom magnetic memory derived from bistability in the orbital configuration of an individual Co atom on semiconducting black phosphorus, yielding two stable and distinct total magnetic moments[29]. Utilizing scanning tunneling microscopy and *ab initio* calculations, we identify and visualize the spatially anisotropic charge densities of each stable valency and detail the effect of the local tip-induced gate potential on the switching behavior between the two. This opens up the possibility of utilizing the orbital degree of freedom for robust single-atom magnetic information storage without requiring spin-sensitive detection, as well as understanding the effect of local gating on the anisotropic charge distribution of a single atomic bit.

**Results**

*Cobalt Deposition and Manipulation*

The result of Co deposition on a BP surface cleaved *in-situ* is shown in Fig. 1a, where the surface illustrates the expected buckled rhombohedral structure. Individual clean Co atoms are identified as a bi-lobed butterfly-like shapes due to the anisotropic extension of their charge density upon adsorption onto BP (see Fig. S1 for larger area images before and after deposition and Fig. S2 for analysis on the presence and influence of hydrogen). As seen in Fig. 1a, two types of bi-lobed Co species are observed (boxed atoms Fig. 1a), related through mirror symmetry along the zig-zag [010] direction, similar to single vacancies in BP[30]. High-resolution analysis of the STM data (Fig. S3) reveals that the bi-lobed species reside on top sites. These species account for approximately ~98% of the as-deposited atoms, indicating favorability toward top-site adsorption during low-temperature ($T \approx 5$ K) deposition; here, the areal density (Fig. 1a) is approximately $0.022 \pm 0.003$ nm$^{-2}$ (see Fig. S1).

Upon current injection with the STM tip above a voltage threshold (Fig. 1d), individual Co atoms can be manipulated from the top site to a hollow site (Fig. 1b and Fig. S4), as confirmed by atomic resolution imaging (Fig. S3). This shift of binding site involves a clear modification to the spatial charge density distribution. Surprisingly, we find that there are two unique and stable shapes of the Co atom within the same hollow site (Fig. 1c, d and Fig. 2), as exhibited by the variation in the charge density. We denote these two states as $J_{H,low}$ and $J_{H,high}$ (index H denotes the atomic site and high/low refers to the size of the magnetic moment shown in Fig. 2e,f). In addition to their unique spatially distributed charge density, $J_{H,high}$ can also be distinguished by its larger apparent height in STM constant-current images ($J_{H,high}$ = 176 ± 8 pm, $J_{H,low}$ = 132 ± 4 pm at $V_s$ = -400 mV). Switching between $J_{H,low}$ and $J_{H,high}$ was achieved via location-dependent current injection (Fig. 1d,e), with $J_{H,high}$ to $J_{H,low}$ at $|V_s| \gtrsim$ 320 mV and $J_{H,low}$ to $J_{H,high}$ at $V_s \gtrsim$ 320 mV. Notably, the switching between different hollow-site states is fully reversible, as shown in Fig S4. However, once a Co atom is manipulated into the hollow site, we were not able to relocate it back into a top site (denoted $J_T$, cf. Fig. 1f). Each of the three atomic configurations remained stable (as probed for measurement times up to 17 hours) until intentionally perturbed. Furthermore, unlike charge switching in single dopants on semiconductors, the atomic state remains fixed after removing the applied bias[16] and the charging lifetime is expected to be very short due to the strong native *p*-doping of the BP crystal[30].

*Ab Initio Calculations and State Identification*

To elucidate the origin of each experimentally observed Co state, we performed density functional theory (DFT) calculations for a Co atom residing on a top (Fig. 2d) and hollow site (Fig. 2e-f) to compare with experimental data (Fig. 2a-c). The calculations were carried out for monolayer BP under the generalized gradient approximation (GGA); to include the effects of local Coulomb interactions in the Co 3*d*-orbital, calculations involving a Hubbard-U correction (GGA+U method) were also performed. Varying the Hubbard-U parameter (Fig. S5), reveals the mutually exclusive stability of two unique states with a critical value at approximately U=3.5 eV, where the state favorability between $J_{H,low}$ and $J_{H,high}$ is inverted. Plotting the spatial distribution of the total charge density from the DFT calculations (Fig. 2d-f), we were able to directly associate each calculation to a corresponding constant-current STM image. The qualitative agreement is excellent and enables us to confirm the experimental binding-site analysis and to roughly approximate the effective screening parameter (U=0-3eV for $J_T$ and $J_{H,low}$, U=4-6 eV for $J_{H,high}$ – see Fig. S5) for the Coulomb repulsion of the Co *3d*-orbital. When including relaxation into the hollow-site calculations, we find that the atomic positions in the surface plane are identical (although out-of-plane distances are different – see Fig 2h,j and the schematic potential diagrams in Fig. 1f); namely the experimental switching from $J_{H,low}$ to $J_{H,high}$ can neither be attributed to a change in binding site nor to different charge configurations.

The use of the Hubbard-U correction allows us to assess distance-dependent screening from the surface within the *3d*-shell of the Co atom. As substrate separation (*d*) is reduced, the more extended *4s*-orbital becomes energetically less favorable due to Pauli repulsion with the BP ligand field, while the increased screening of the *3d*-orbital increases its energetic favorability by decreasing Coulomb repulsion in the system. The resulting occupation of the Co *4s* ($n_s$) and *3d* ($n_d$) orbitals is given in Fig. 2 for each of the states (resolved into the *3d*-subshells in table S1). We find from these calculations, that the relaxation (Δ*d*) from $J_{H,low}$ to $J_{H,high}$ (Fig. 2h,j) is accompanied by a redistribution of the 4*s*- and 3*d*-orbital populations (for further detail see Fig. S6, S7, and S8). As expected when modifying metal *3d*-orbital occupancy, the total magnetic moment also changes between 1.00 $\mu_B$ for $J_{H,low}$ and 2.34 $\mu_B$ for $J_{H,high}$. Furthermore, calculations of the magnetic anisotropy indicate that the easy axis also changes from in-plane ($J_{H,low}$) to out-of-plane ($J_{H,high}$) (table S2). This suggests that the magnetic anisotropy of Co can be controlled electrically in this system. We note that similar orbital behavior has been predicted for transition metal atoms on graphene[29,31,32], where multiple states (different *d*) were analogously predicted due to the reorganization of the orbital occupancies. Quantum chemistry calculations for Co on graphene further indicated that the energy barrier between states could reach nearly 300 mV[29], which might explain the remarkable stability of the states observed here. This also indicates that using the orbital degree of freedom may be much more robust compared to using solely the bistability of the spin ground states.

*Tip-Induced Local Gating*

In order to elucidate the valency switching mechanism, we studied the influence of tip-induced band bending (TIBB) on the Co states. Due to insufficient screening from charge carriers, the applied potential between tip and sample locally influences the energy of semiconductor bands; if a impurity level, shifted with the material bands, passes through the Fermi level ($E_F$), it will undergo an observable charging/discharging event in STM/STS[33]. Such charging events resulting from TIBB can be distinguished by peaks in d$I$/d$V$ whose location and intensity are strongly sensitive to the stabilization parameters (Fig. S9) and tip location (Fig. S10) [33,34]. While all states demonstrate ionization events, we limit our focus in this work to the $J_{H,low}$ and $J_{H,high}$ states. A representative d$I$/d$V$ spectrum for $J_{H,low}$ (Fig. 2m) clearly shows a strong peak at approximately 280 mV, while the primary peak for $J_{H,high}$ (Fig. 2n) is seen at 420 mV. In conjunction with the spectroscopic mapping (see below), the shaded regions (labeled $q^+$) are identified as bias ranges where the Co species have been ionized. At biases greater than these thresholds, the atoms are non-locally ionized via the tip-induced potential along the BP surface.

To gain a more complete picture of this local surface potential, we used constant-height imaging to map out the spatial dependence of the ionization as a function of bias voltage (Fig. 3a,b)[34]. The size of the isotropic disk (stepwise increase in current around the Co, or so-called charging ring when imaged in d$I$/d$V$ maps - Fig. S10 and Fig. S11), scales similarly for both states with bias according to hyperbolic contours of constant TIBB (Fig. 3c). This indicates equivalent screening from the BP for each Co configuration. Furthermore, the trend of the effective ring radii ($r_{eff} = L/2\pi$, where $L$ is the ring circumference) with applied bias (see also Fig. S10) indicates a flat-band condition of $V_{FB} < -300$ mV (Fig. 3d). Such a condition is achievable with a tip work function of 4.0-4.1 eV. Identifying the flat-band condition and the ring-radius dependence on bias indicates that the ionization events are caused by the upward bending of states below $E_F$ (see Fig. 3d). Theoretical calculations for both configurations reveal non-zero density of states below $E_F$; however, $J_{H,low}$ clearly has a strong 3$d$-orbital peak in the DOS between $E_F$ and the valence band edge ($E_v$). Consistently smaller radii for $J_{H,high}$ compared to $J_{H,low}$ indicate that larger TIBB is needed to ionize $J_{H,high}$ (near 400 mV); thus this state must lie farther from $E_F$ than the ionized state of $J_{H,low}$ (Fig. S10h).

*Switching Dynamics and Mechanism*

Upon gating the Co into the charged regimes ($q^+$ regions in Fig. 2m,n) with the STM tip, a discrete, bistable conductance signal, or so-called telegraph noise, is measured on the Co atoms (Fig. 4a). The bistable states are correlated to the $J_{H,low}$ (dark blue) and $J_{H,high}$ (light blue) configurations of Co via independent constant-current measurements at lower negative biases (-400 mV < $V_s$ < -200 mV), which do not perturb the respective states. The ability to read and write both Co orbital configurations confirms its utility as a binary memory; thus we denote $J_{H,low}$ as state "0" and $J_{H,high}$ as state "1". To further probe the stability and favorability of these states, we studied the bias and spatial dependencies of the telegraph noise. We define τ as the residence time for the given state, as derived in ref. 35. Fig. 4b illustrates the effect of bias at constant height. Maintaining a constant tip-sample separation, while smoothly varying the bias, results in a proportional change in the TIBB at the surface. In this manner, the influence of TIBB on the state stability can be directly measured without artifacts introduced by changing the tip-sample distance. As seen in the top panel of Fig. 4b, at lower biases $τ_{H,low}$ and $τ_{H,high}$ are nearly equivalent and strongly decay with increasing TIBB, resulting from the increased energy and flux of the tunneling electrons. However, above a threshold voltage $V_s = V^* \approx 540$ mV the state-dependent lifetimes diverge from each other, leading to a large state favorability or what we define as asymmetry ($A = (\frac{\tau_{H,high} - \tau_{H,low}}{\tau_{H,high} + \tau_{H,low}})$) (lower panel Fig. 4b), whereby $τ_{H,high}$ is strongly diminished. This divergence indicates that given sufficient gating (above the critical bias threshold $V^*$), there is a strong energetic favorability in the decay mechanism from $J_{H,high}$ to $J_{H,low}$ (Fig. 4c). Reexamining the charging ring data from Fig. 3c, we see that this critical bias corresponds to a $J_{H,high}$ critical charging ring radius ($r^*$) of approximately 2 nm. We also note here that the onset of telegraph switching occurs after the ring radius for $J_{H,low}$ exceeds $r^*$. These observations suggest that a minimum gate potential (measured as a ring radius $r^*$) is required to achieve efficient switching for both states. This threshold is likely related to the extension of the ionized Co charge density, which can span 2-4 nm (see Fig. 5a), as the screening for both states is nearly identical. Based on these observations, we sketch the qualitative energy diagram for the subcritical ($r_{eff} < r^*$, Fig. 4c left panel) and supercritical ($r_{eff} > r^*$, right panel) regimes; significantly, the only observed barrier modification is the one between ionized species ($E^*$).

**Discussion**

Finally, to understand the connection between the TIBB and the atomic charge density, including the subsequent impact on the switching behavior, we studied the sensitivity of the telegraph noise to the precise tip-gate position (Fig. 5). The upper panel of Fig. 5d shows the mean lifetime ($\tau_M = (\tau_{H,high} + \tau_{H,low})/2$) as a function of position across the Co atom (at constant height) along two orthogonal directions (shown with darker gray ([010]) and lighter gray ([100]) arrows in Fig. 5a-c). The curves reveal contrasting results: along [100], the switching persists anisotropically to a distance of ~3.5 nm from the atom, while the switching rate decays symmetrically and significantly faster along [010] (dark gray). The differences are further reflected in the asymmetry of the states along the two distinct directions, where the zig-zag direction shows almost complete suppression of $\tau_{H,high}$ (inset, Fig. 5d). The spatial dependencies of Fig. 5d closely match the anisotropy of the $J_{H,low}$ ionized charge density (Fig. 5a), pointing to a switching mechanism based on the overlap between TIBB and the spatial extent of the charge density. We note that this switching anisotropy can be utilized to reach almost 100% directed switching probability, which is an important requirement for controlled writing of the single-atom memory.

In conclusion, we show that the total magnetic moment, i.e. the valency, of an individual Co atom can be controllably switched electrically, and is an extremely robust means to store information. The discrete states result from a shift in the relative orbital population between *4s* and *3d* states of the Co atom that is driven by a change of the effective Coulomb screening between the Co atom and the BP surface. Small differences in the charge and electric dipole nature between the two states could be further elucidated with atomic-scale Kelvin probe experiments to qualitatively probe the valency of each state[36]. Utilizing bistability in the orbital degree of freedom presents many advantages compared to single magnetic atom spin-based computing[3,35]. First, these states can be both read and written electrically without spin sensitivity. Second, the interplay between the spin and orbital moment may also possess a new route toward multi-bit registers. Third, this work also sheds light on the effect of orbital switching by local gating of individual dopants with an anisotropic charge density. Unlike previous studies of single hydrogenic impurities in semiconductors, this work reveals the significance of both TIBB and wavefunction anisotropy in the stability of a two-state system. Our calculations also show that the orientation of the magnetic anisotropy and its amplitude are significantly modified for each of the bistable valencies, indicating a method of electrically controlling the magnetic anisotropy. This motivates future experiments based on spin-resolved STM and inelastic tunneling spectroscopy, which may reveal the nature of the magnetic anisotropy[15] as well as the spin lifetimes of each Co valency[7]. Finally, the energy separation between orbitals can be significantly larger than magnetic anisotropy or Zeeman energies, potentially making it viable for room temperature application. As single Co atoms were also observed after annealing the sample to room temperature (Fig. S12), this may open the possibility for realistic higher temperature applications; however, subsequent experiments revealing the energy barriers between the two states will prove pivotal to ascertaining the potential value of such a system for room temperature information storage.

**Acknowledgements**

The authors would like to acknowledge scientific discussions with Nadine Hauptmann, Jill Miwa, and Michael Flatté. B.K. and A.A.K. acknowledge financial support from the Emmy Noether Program (KH324/1-1) via the Deutsche Forschungsgemeinschaft, and The Netherlands Organization for Scientific Research (NWO). B.K. and A.A.K. also acknowledge the VIDI project: 'Manipulating the interplay between superconductivity and chiral magnetism at the single atom level' with project number 680-47-534 which is financed by NWO. A.N.R. acknowledges support from the Russian Science Foundation, Grant 17-72-20041. This work was supported by VILLUM FONDEN via the Centre of Excellence for Dirac Materials (Grant No. 11744). B.K. and A.A.K. acknowledges funding from the European Union's Horizon 2020 research and innovation programme under grant agreement No. 751437.

**Author Contributions**

B.K. and W.M.J.v.W. performed the experiments and data analysis with the help of A.A.K. and DW. A.N.R. and M.I.K. performed the calculations and analysis of the theoretical data. A.A.K. and B.K. designed the experiments. B.K., A.A.K., D.W., and A.N.R. wrote the manuscript. All authors provided input for the manuscript and discussion.

**Methods**

*Scanning Tunneling Microscopy/Spectroscopy*

STM/STS measurements were performed in ultrahigh vacuum (< $1 \cdot 10^{-10}$ mbar) on an Omicron low-temperature STM with a base temperature of 4.4 K, with the bias applied to the sample.

Electrochemically etched W tips were used for measurements; the tips were treated *in situ* by electron bombardment, field emission, as well as dipped and characterized on a clean Au surface. Scanning tunneling spectroscopy was collected using a lock-in technique to directly measure d*I*/d*V*, a modulation frequency of $f_{mod}$ = 4.2 kHz and amplitude of $V_{mod}$ = 2-6 mV were applied to the bias signal. Black phosphorus crystals were provided by HQ graphene and subsequently stored in vacuum (< 1 × $10^{-8}$ mbar) at a temperature less than 25 °C. The crystals were cleaved under ultrahigh vacuum conditions at pressures below 2·$10^{-10}$ mbar, and immediately transferred to the microscope for *in-situ* characterization. Cobalt was evaporated directly into the STM chamber with $T_{STM}$ < 5 K for the entire duration of the dosing procedure.

*Theoretical Calculations*

DFT calculations were carried out using the projected augmented-wave method (PAW)[37] as implemented in the Vienna *ab initio* simulation package (VASP)[38,39]. Exchange and correlation effects were taken into account within the spin-polarized generalized gradient approximation (GGA) in the parametrization of Perdew-Burke-Ernzerhof (PBE)[40]. Additional Hubbard-U correction was applied to the 3d shell of Co within the GGA+U method[41] in order to capture the effect of the distance-dependent Coulomb screening. An energy cutoff of 300 eV for the plane-wave basis and the convergence threshold of $10^{-6}$ eV were used in the self-consistent solution of the Kohn-Sham equations, which checked to be sufficient to obtain numerical accuracy. Pseudopotentials were taken to include 3*s* and 3*p* valence electrons for P atom, as well as 3*s*, 3*p,* and 3*d* valence electrons for Co atom. BP surface was modeled in the slab geometry by a single BP layer with dimensions (3*a* × 4*b*) ≈ (13.1 × 13.3) Å with atomic positions fixed to the experimental parameters of bulk BP[42]. Vertical separation between the layers was set to 20 Å. The Brillouin zone was sampled in two-dimensions by a uniform distribution of **k**-points on a (8 × 8) mesh. The position of Co atom was relaxed considering different surface sites (top and hollow) as starting points. We checked that the inclusion of two additional BP layers in the slab does not significantly affect the results presented in the main part. The primary difference between the single- and three-layer slabs is the reduction of a gap between the valence and conduction BP states. The behavior of Co atom remains virtually unchanged including the adsorption distances, charge density distribution, and magnetic moments. The charge density distributions shown on Fig. 2d-f were obtained by averaging the total charge density over the energy interval of ~0.3 eV in the valence band edge. The projection of the electronic bands on specific atomic states was done using the formalism of maximally localized Wannier functions[43] implemented in the wannier90 package[44].

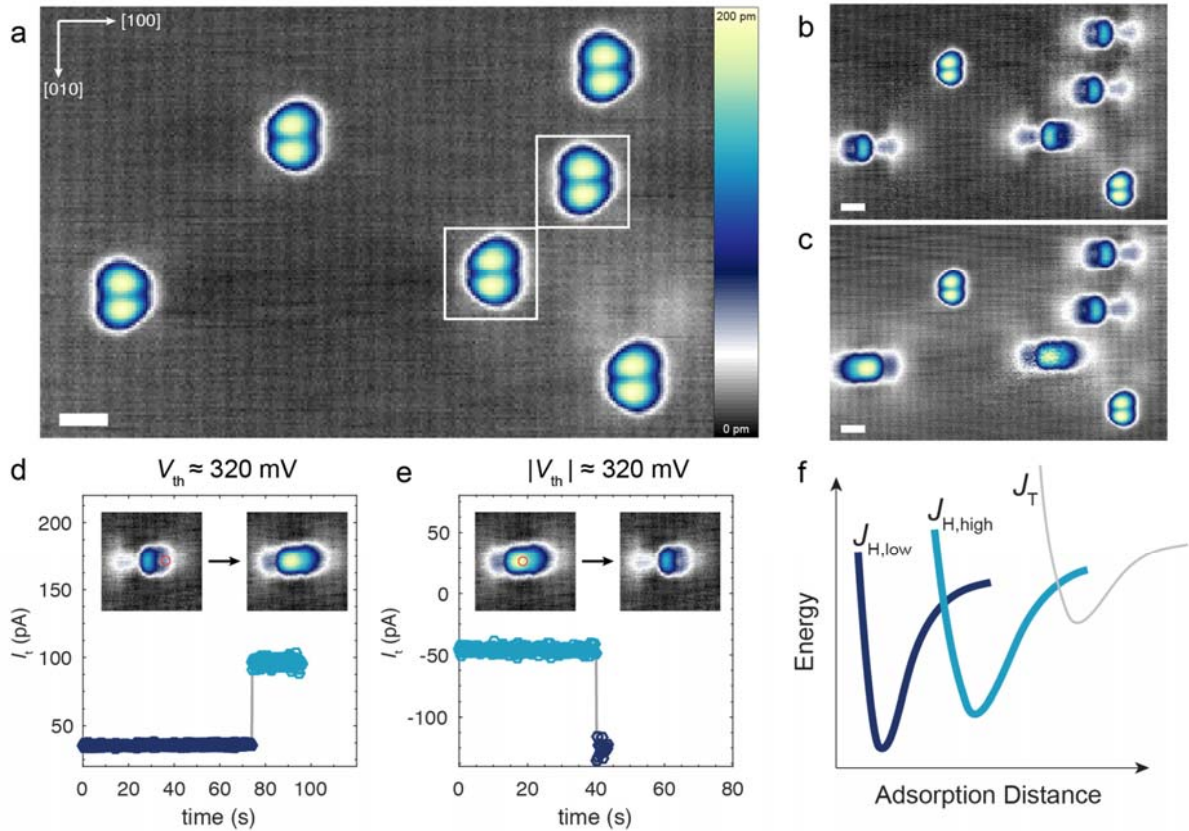

Fig. 1: Adsorption and switching of Co on BP. (a) Six Co species on BP as deposited at $T < 5$ K ($V_s$ = -400 mV, $I_t$ = 20 pA, scale bar = 1 nm). Boxed atoms show species related through mirror plane along [010]. (b) Four atoms from (a) have been switched into $J_{H,low}$ ($V_s$ = -400 mV, $I_t$ = 20 pA, scale bar = 1 nm). (c) Two atoms from (b) have been switched into $J_{H,high}$ ($V_s$ = -400 mV, $I_t$ = 20 pA, scale bar = 1 nm). (d) Switching characteristics from $J_{H,low}$ to $J_{H,high}$ with $V_s$ = 420 mV and (e) $J_{H,high}$ to $J_{H,low}$ with $V_s$ = -680 mV. Approximate threshold biases for switching ($V_{th}$) are noted. Orange circles indicate the tip position during the switching sequence. The inset images showing before and after configurations are 4 nm x 4 nm in size. (f) Schematic representation of adsorption energy curves for Co species on BP.

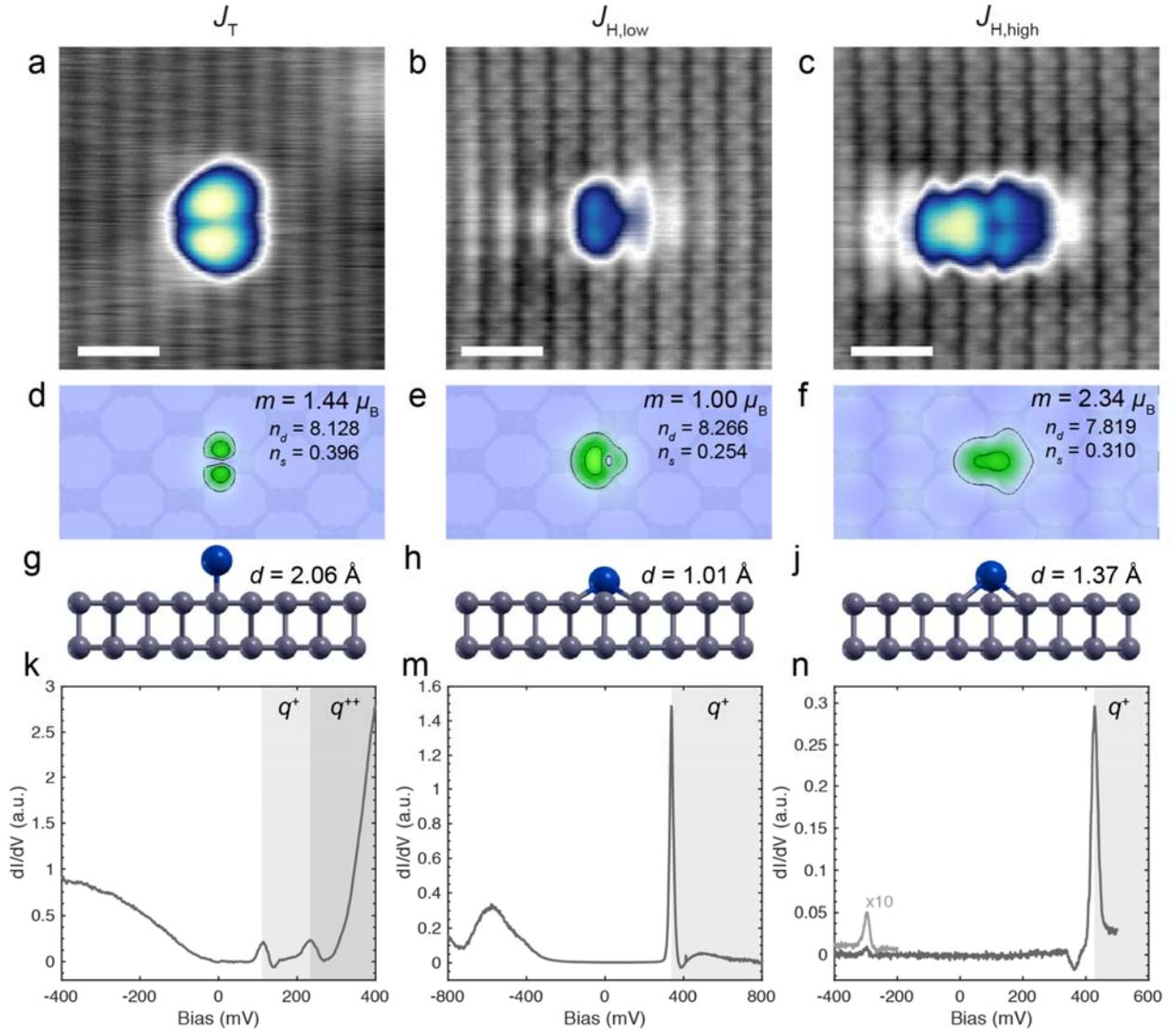

Fig. 2: Ground states of Co atoms. High-resolution image of Co in (a) $J_T$ ($V_s$ = -400 mV, $I_t$ = 200 pA, scale bar = 1 nm), (b) $J_{H,low}$ ($V_s$ = -60 mV, $I_t$ = 200 pA, scale bar = 1 nm), and (c) $J_{H,high}$ ($V_s$ = -60 mV, $I_t$ = 200 pA, scale bar = 1 nm) configurations with same color scale as Fig. 1. DFT calculations of charge density distributions, including magnetic moment ($m$), $n_d$, and $n_s$, for (d) Co on a top site, (e) Co in a hollow site, and (f) Co in a hollow site with U = 4 eV. (g-j) Schematics of relaxed atomic adsorption geometries with out-of-plane distance ($d$) noted. (k-n) d$I$/d$V$ spectra taken on each atom.

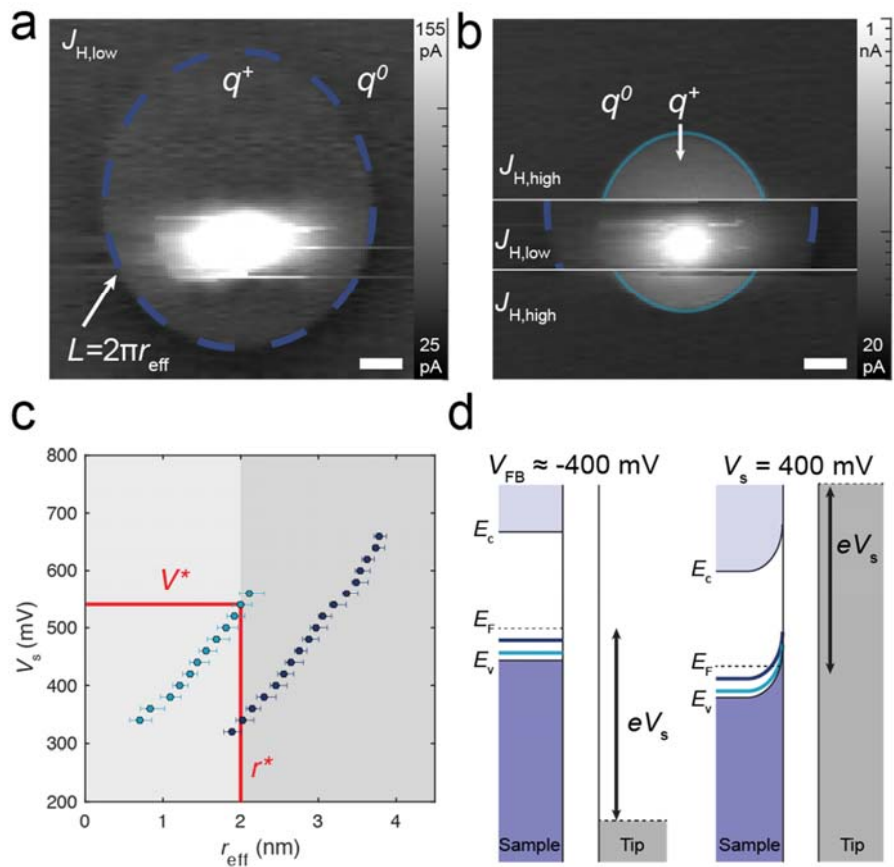

Fig. 3: Ionization of Co. (a) Constant-height current map of $J_{H,low}$ (setpoint conditions: $V_s$ = 500 mV, $I_t$ = 40 pA), showing isotropic charging state (highlighted with dashed blue ring) as a disk in the current map (logarithmic color scale, scale bar = 1 nm). Charged ($q^+$) and uncharged ($q^0$) regions are denoted. (b) Constant height map of same Co atom initialized into $J_{H,high}$ state (setpoint conditions: $V_s$ = 500 mV, $I_t$ = 40 pA, scale bar = 1 nm, logarithmic color scale). The atom switches into $J_{H,low}$ for a small section in the middle of the image. (c) Charging ring effective radius ($r_{eff} = L/2\pi$) as a function of bias for $J_{H,low}$ (dark blue), and $J_{H,high}$ (light blue). Red lines show threshold bias ($V^*$) determined from switching data and corresponding critical radius ($r^*$) for $J_{H,low}$ and $J_{H,high}$. (d) Schematics of proposed flat-band condition ($V_{FB} \approx -0.4$ eV) and band bending at selected bias near (a),(b).

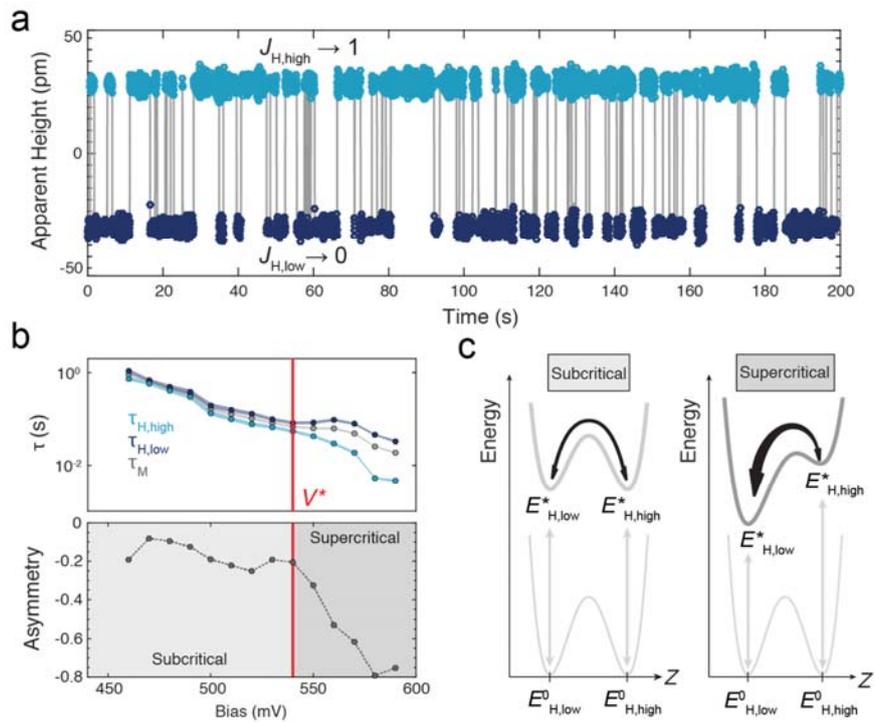

Fig. 4: Orbital memory. (a) Two-state conductance signal with $J_{H,high}$ (light blue) showing high conductance and $J_{H,low}$ (dark blue) showing low conductance. (b) State lifetime and asymmetry as a function of tip bias with tip height held constant. Linewidth of lifetime curves indicates measurement error. $V^*$ highlights critical threshold where TIBB causes lifetime depression for $J_{H,high}$. (c) Energy schematics illustrating excitation/ionization ($E^0$ to $E^*$) and switching mechanism for orbital memory. (Left) Sub-threshold behavior with nearly equivalent lifetimes (symmetric upper well structure) and (right) supercritical behavior with strongly diminished $J_{H,high}$ lifetime (imbalanced upper well structure).

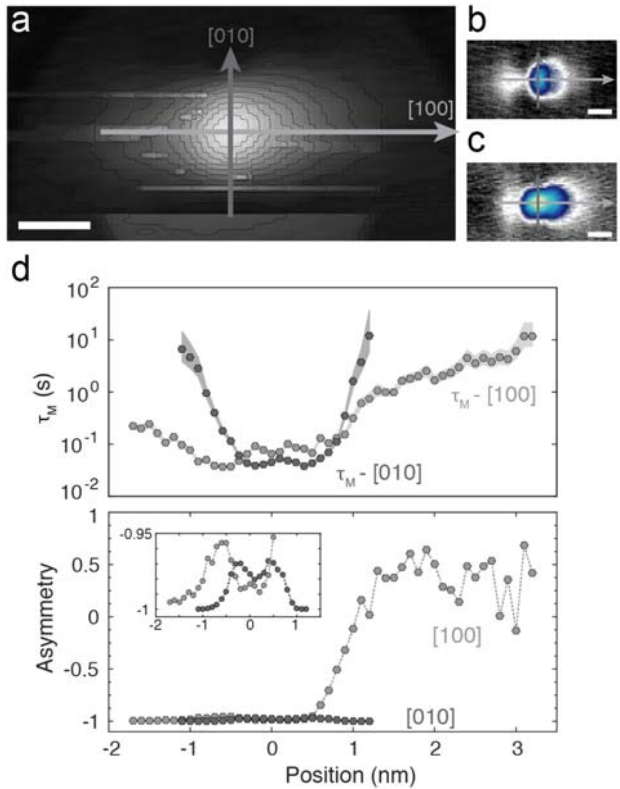

Fig. 5: Anisotropic Gating. (a) Constant-height current map showing Co charge density at $V_s$ = 500 mV (setpoint conditions: $V_s$ = 500 mV, $I_t$ = 40 pA, scale bar = 1 nm). Dark gray and light gray arrows show measurement positions for (d) along [010] and [100], respectively. (b) Measurement locations shown on constant-current image of $J_{H,low}$ ($V_s$ = -400 mV, $I_t$ = 20 pA, scale bar = 1 nm). (c) Measurement locations with reference to $J_{H,high}$ ($V_s$ = -400 mV, $I_t$ = 20 pA, scale bar = 1 nm). (d) Spatially resolved constant height measurements (setpoint conditions: $V_s$ = 500 mV, $I_t$ = 40 pA). Mean lifetime (top) and state asymmetry (bottom) as a function of position along (light gray) [100] and (dark gray) [010] directions. (Inset, bottom panel) Rescaled plot of asymmetry data from positions -2 nm to 1 nm. The constant height image in (a) was taken with precisely the same tip height and bias to show the length scale of the ionized atomic wavefunction (atom is primarily in $J_{H,low}$). Position 0 nm corresponds to the center of the Co atom. Data was collected with $V_s$ = 500 mV.

Supplementary Information

# An orbitally derived single-atom magnetic memory

Brian Kiraly[1], Alexander N. Rudenko[2,1,3], Werner M.J. van Weerdenburg[1], Daniel Wegner[1], Mikhail I. Katsnelson[1], Alexander A. Khajetoorians[1]

1. Institute for Molecules and Materials, Radboud University, Nijmegen, The Netherlands
2. School of Physics and Technology, Wuhan University, Wuhan 430072, China
3. Theoretical Physics and Applied Mathematics Department, Ural Federal University, 620002 Ekaterinburg, Russia

Table of Contents

**Cobalt deposition, identification, and manipulation on black phosphorus**

Figure S1 shows constant-current scanning tunneling microscopy (STM) images of the black phosphorus (BP) surface before (Fig. S1a) and after (Fig. S1b) deposition of Co atoms. To avoid Co deposition on the STM tip, the tip was fully retracted during the dose; thus the images do not reflect the exact same sample area. A characteristic vacancy is seen in the constant-current image in Fig. S1a, with a similar vacancy in Fig. S1b for comparison; both images are shown with the same color scale for apparent height. As described in the main text and seen in Fig. S1b, approximately 98% of the cobalt atoms initially appear as bi-lobed species (labeled $J_T$) with an apparent height of approximately 180 pm. The coverage shown in Fig. S1b is 0.026 nm$^{-2}$. For detailed studies of the switching dynamics, the coverage was reduced by approximately an order of magnitude to ensure minimal influence from non-local interactions.

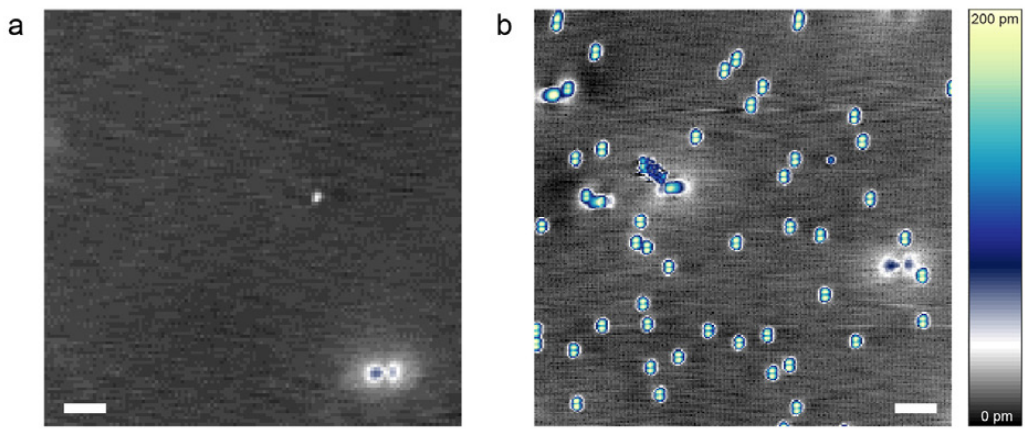

**Supplementary Figure 1. Deposition of Co on BP.** (a) Black phosphorus crystal after cleaving and inserting into STM ($V_s$ = -400 mV, $I_t$ = 20 pA, scale bar = 4 nm). (b) Surface immediately after deposition of Co atoms at $T$ = 5 K ($V_s$ = -400 mV, $I_t$ = 20 pA, scale bar = 4 nm).

It is well known that hydrogen can have a significant impact on the properties of atoms[1-3]. For Co on BP, we show that the addition of a hydrogen atom or atoms to $J_T$ (denoted CoH$_x$, where x is an unknown integer) causes (left panel, Fig. S2a): (1) a reduction of the apparent height (CoH$_x$ ≈ 140 pm) and (2) a modification to the spatial charge density. The species shown in the leftmost panel of Fig. S2a is hypothesized to be $CoH_x$ because it appears slowly over time (Fig. S2b-e) and can be irreversibly modified with tip-induced manipulation into $J_T$. This is currently the only species displaying such behavior, consistent with previous reports describing hydrogen desorption from adatoms[2,4]. As shown in Fig. S2a, after removing the hydrogen, the atomic switching behavior ($J_T$ to $J_{H,low}$ and $J_{H,low}$ to $J_{H,high}$) is as expected, signaling the return of the Co atom to its pristine state. The density of such $CoH_x$ species on a single sample (held at $T = 4$ K) over the period of 40+ days is shown in Fig. S2e. In order to minimize the influence of hydrogen, two procedures were used: (1) the cryostat was regularly warmed to $T > 40$ K and the system pumped with a turbomolecular pump to remove adsorbed hydrogen from the cryostat, and (2) samples were generally studied for less than 10 days to minimize hydrogen contamination. Control spectra were routinely taken on the BP to ensure no spurious features were observed. Finally, the reported spectroscopic features in Fig. 2 and Fig. S9 were observed with multiple unique tips and did not vary with time.

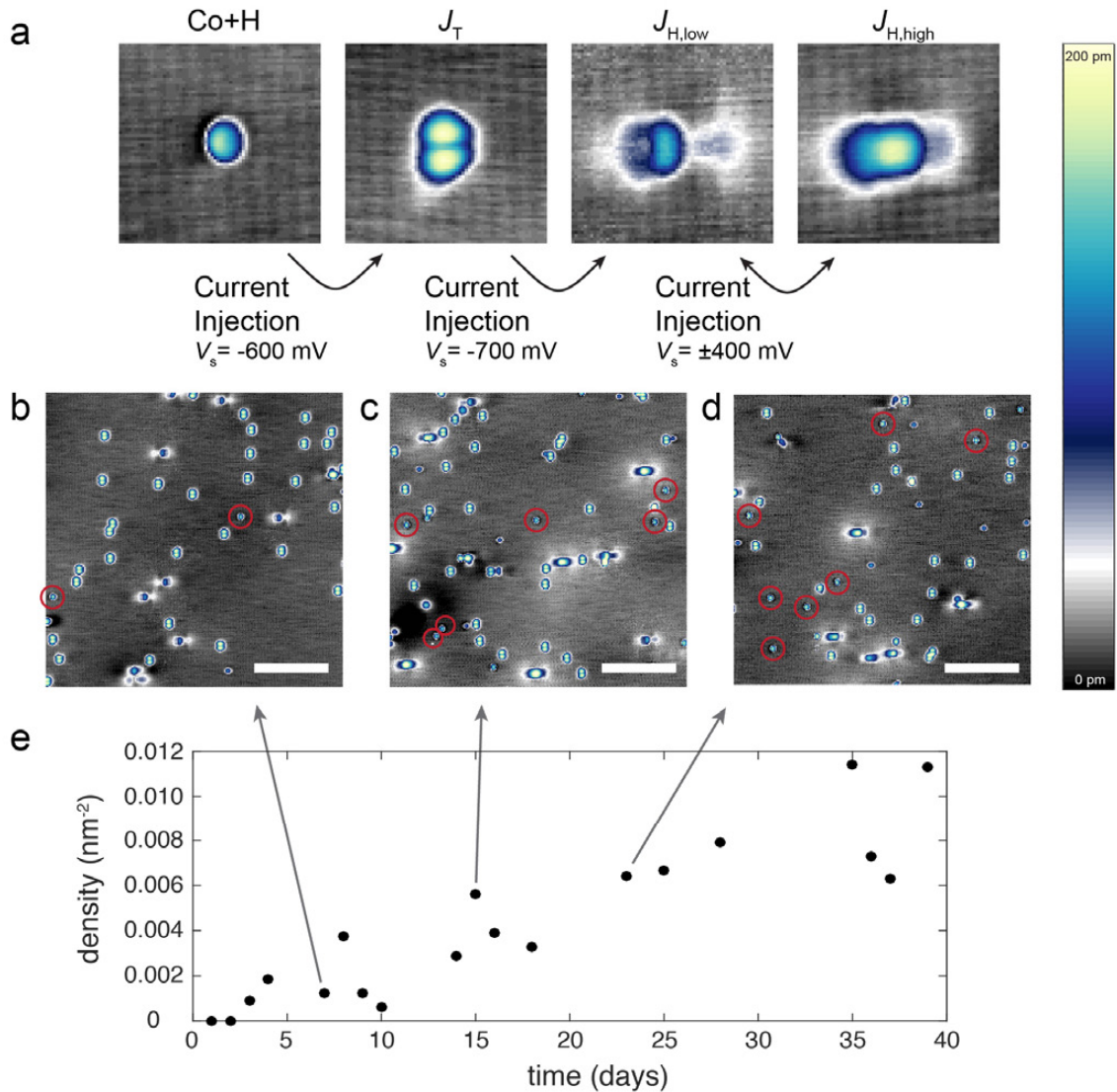

**Supplementary Figure 2. Characterization of hydrogen contamination.** (a) Hydrogenated Co species (red circles in (b), (c), and (d)) tentatively identified by the ability to manipulate them back into the $J_T$ state (second panel). Expected switching behavior ($J_{H,low}$ third panel and $J_{H,high}$ fourth panel) indicates a return of the Co to a pristine state. All images are 4 nm x 4 nm ($V_s$ = -400 mV, $I_t$ = 20 pA). (b), (c) and (d) show STM images of the sample at 7, 15, and 23 days after deposition, respectively ($V_s$ = -400 mV, $I_t$ = 20 pA, scale bar = 4 nm). (e) Density of Co+H species other than $J_T$, $J_{H,low}$, and $J_{H,high}$ on a single sample held at 4 K for over 40 days.

Binding site analysis was done via atomic resolution imaging at sample biases near the valence band edge (generally -100 mV < $V_s$ < -30 mV) to clearly resolve all P atoms in the upper zig-zag row of the BP surface layer. Unfortunately, at the optimal conditions for BP atomic resolution ($V_s$ = -30 mV, $I_t$ = 20 pA), the $J_T$ state was highly unstable, either shifting binding sites or switching into $J_{H,low}$. Thus, a sequence of images was used to determine the $J_T$ binding site (Fig. S3a-c). First, an image was taken at $V_s$ = -400 mV to determine the atomic position (color image Fig. S3a). The second, partial image was then taken at $V_s$ = -30 mV to determine the BP lattice sites (overlaid image Fig. S3a). Finally, a third image was taken at $V_s$ = -400 mV to recheck the atomic position. Using the red contours of constant charge density (Fig. S3b,c), the geometric center of the atom was determined to reside on a phosphorus top site, as indicated in Fig. S3c. Atomic resolution of the $J_{H,low}$ and $J_{H,high}$ states was acquired simultaneously with the BP lattice (Fig. S3d,g) despite occasional switching events (Fig. S3d). Analysis of the constant charge-density contours (Fig. S3e,h) shows that both species' geometric center resides within a hollow site. Furthermore, the binding site data (Fig. S3d-j) confirms that both the $J_{H,low}$ and $J_{H,high}$ states for a single atom reside in the same hollow site. Finally, the contours of constant charge density (Fig. S3c,f,j) for all three states agree well with the DFT charge density predictions given in Fig. 2k-n, further confirming the binding site assignment.

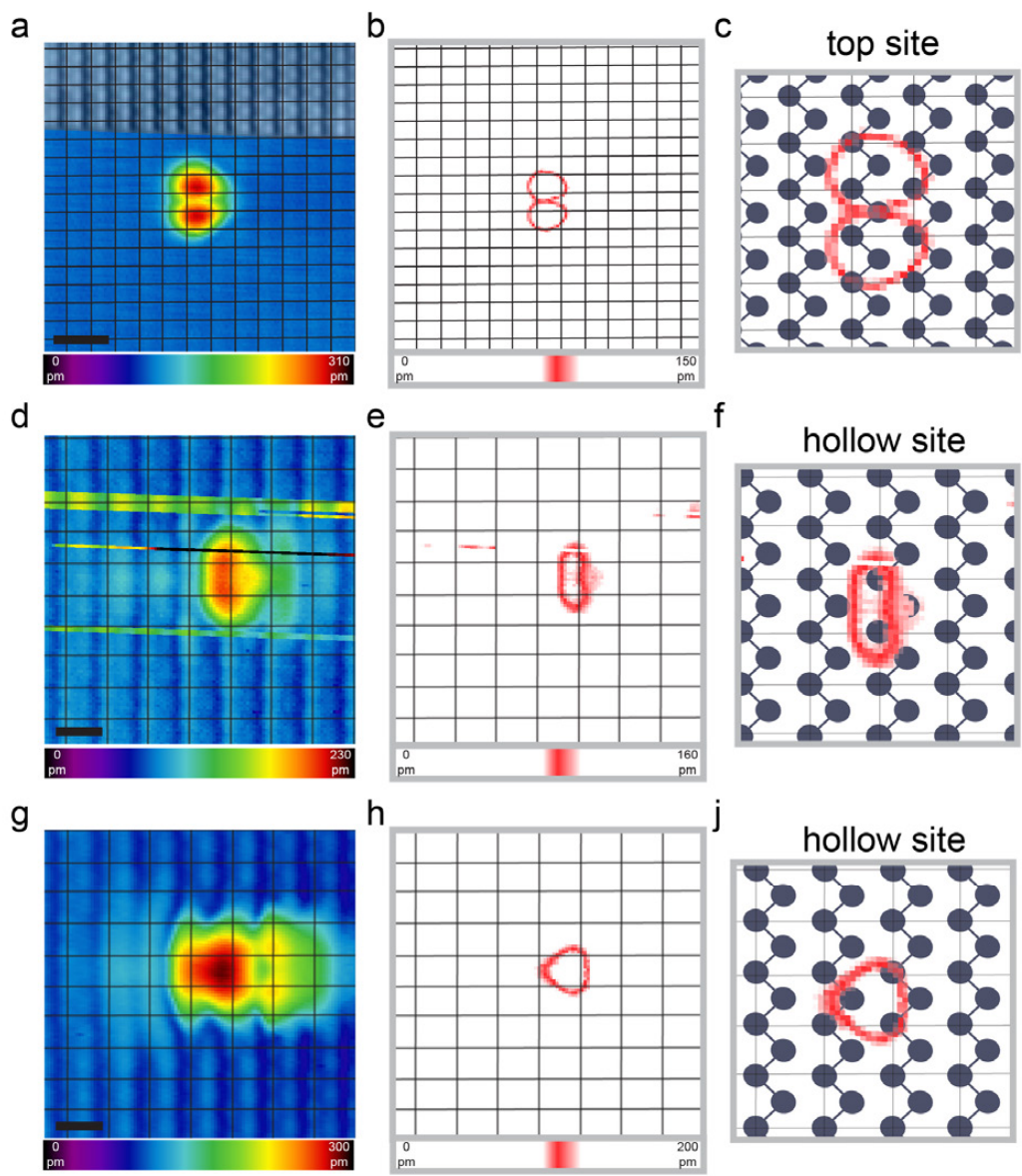

**Supplementary Figure 3. Atomic resolution analysis of binding site.** (a) Top-site atom with sublattice-resolved image overlaid on an adatom image (adatom image: $V_s$ = -400 mV, $I_t$ = 20 pA, scale bar = 1 nm, sublattice image: $V_s$ = -30 mV, $I_t$ = 20 pA). The black phosphorus conventional unit cell (black – reflecting the periodicity of one of the two sublattices) as determined by the sublattice-resolved image is overlaid on the STM data. (b) STM data from (a) is replotted to select a single range of apparent height (charge density) to clearly denote the location of the Co on the BP lattice. (c) Binding site analysis using the red contour of constant charge density to determine the geometric center of the atom. The actual BP lattice has been included for clarity based on the unit cell from (a,b). (d) Atomic resolution image of $J_{H,low}$ state with sublattice resolution of the BP ($V_s$ = -30 mV, $I_t$ = 20 pA, scale bar = 0.5 nm). Again, the BP conventional unit cell is overlaid in black. (e) Contour of constant charge density plotted with the BP lattice. (f) Hollow binding site designation based on the constant charge density contour. (g) Atomic resolution image of $J_{H,high}$ state with sublattice resolution ($V_s$ = -30 mV, $I_t$ = 20 pA, scale bar = 0.5 nm). Black phosphorus lattice overlaid in black. (h) Contour of constant charge density plotted with the BP lattice. (j) Hollow binding site designation based on the constant charge density contour. All images have been corrected for a 2° offset between the x and y directions.

A step-wise switching sequence, showing both the localization and reversibility of the switching, is given in Fig. S4. Figure S4b-e show the switching of atoms 1-4 (Fig. S4a), respectively, from the $J_T$ state into the $J_{H,low}$ state. The isolated manipulation of atoms 2-4 (Fig. S4c-e) despite their ~ 2 nm separation, demonstrates the confinement of the tip-induced perturbation responsible for the switching. Figure S4f,g further shows the switching of atoms 1 and 2 from the $J_{H,low}$ state into the $J_{H,high}$ state. To demonstrate the reversibility of the $J_{H,low}$ to $J_{H,high}$ transition, Fig. S4h,j show the reverse switch from the $J_{H,high}$ state back into the $J_{H,low}$ state. As seen in the telegraph data from Fig. 4, reversible switching between $J_{H,high}$ and $J_{H,low}$ was observed 1000's of times without modification to the constituent states.

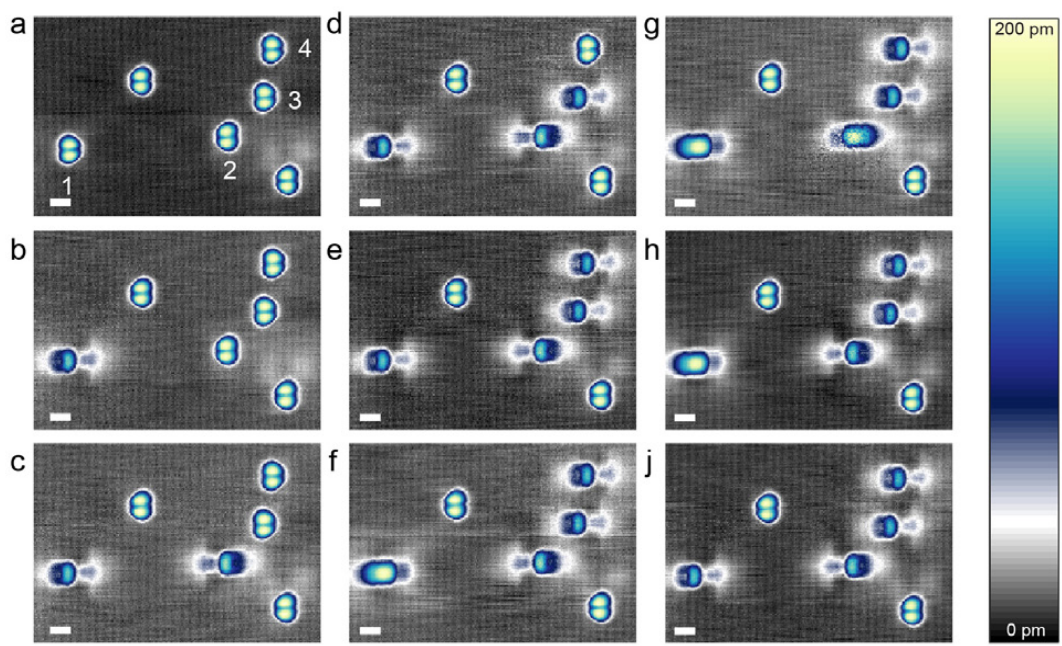

**Supplementary Figure 4. Full switching sequence.** (a) Initial state for all six species. (b-e) Switching atoms 1-4 from $J_T$ site to $J_{H,low}$. (f,g) Switching atoms 1-2 from $J_{H,low}$ into $J_{H,high}$. (h,j) Switching atoms 1-2 from $J_{H,high}$ into $J_{H,low}$. All images: $V_s$ = -400 mV, $I_t$ = 20 pA, scale bar = 1 nm.

## *Ab-Initio* calculations for cobalt/black phosphorus density of states and band structure

Figure S5 shows total energies of the $J_{H,low}$ and $J_{H,high}$ states calculated as a function of the Hubbard-U parameter. One can see that for U < 3.5 eV the $J_{H,low}$ state is energetically favorable, whereas at U > 3.5 eV the $J_{H,high}$ state becomes more stable. The corresponding charge distribution averaged over the valence band edge (as in Fig. 2d-f) are shown at the bottom of Fig. S5. In these panels, it is clear that the shape of the charge distribution is different for the two orbital states, irrespective of the exact value of the Hubbard-U parameter for each state. In this theoretical analysis, the Hubbard-U is a phenomenological parameter related to the Coulomb repulsion. To adequately describe two distinct orbital configuration of Co on BP it is therefore sufficient to consider U=0-3 eV for $J_{H,low}$ and U=4-6 eV for $J_{H,high}$. For comparison with experimental STM images, we use U=0 and U=4 eV for the $J_{H,low}$ and $J_{H,high}$ states, respectively.

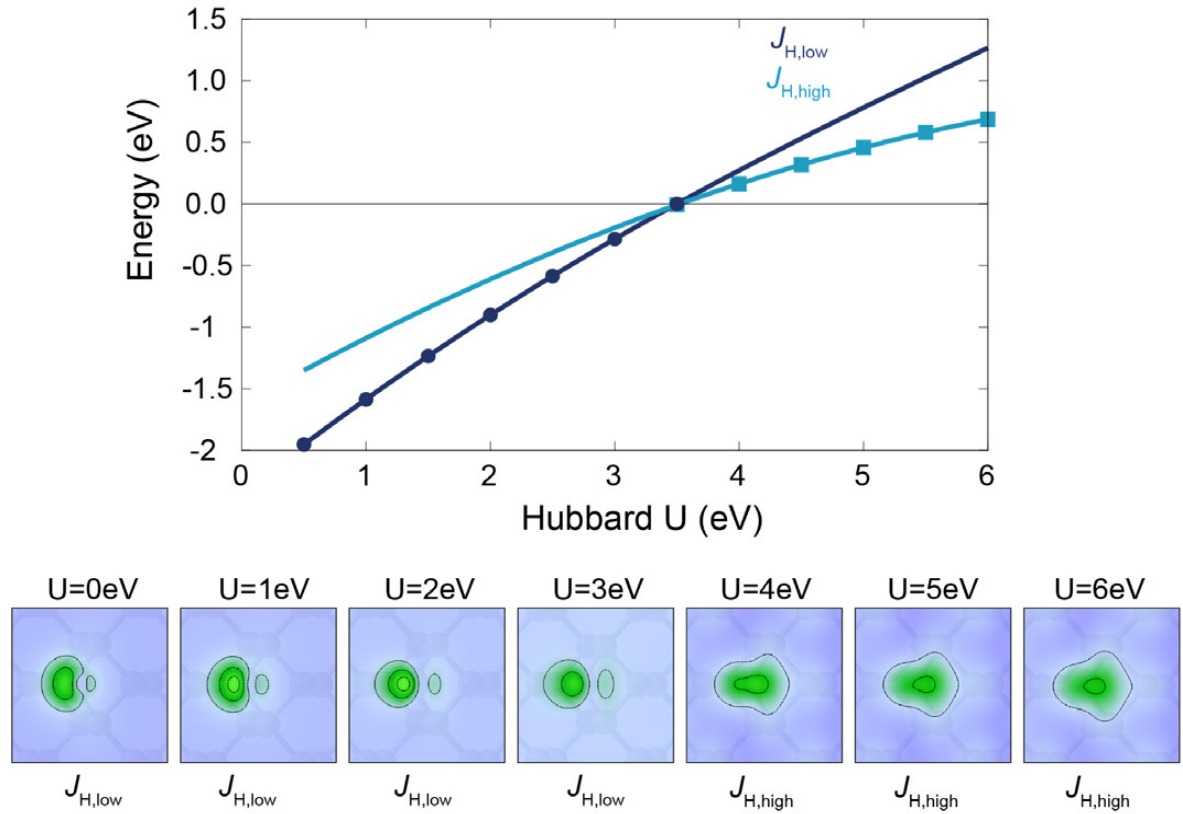

**Supplementary Figure 5. Calculations with varying Hubbard-U correction.** (upper) Plot showing total energy (markers) of $J_{H,low}$ state (dark blue) for U < 3.5 eV and $J_{H,high}$ (light blue) for U > 3.5 eV. Solid lines show fits to the markers. The state favorability inverts at U = 3.5 eV, where both states are stable and nearly degenerate in energy. (lower) Plots of calculated charge density distributions calculated for varying Hubbard-U parameters.

In Fig. S6, we show spin-resolved density of states (DOS) calculated at the density functional theory (DFT) level for the three states of cobalt atom on BP discussed in the paper. For the $J_T$ and $J_{H,low}$ states, sharp peaks in the vicinity of the Fermi level are visible, originating from cobalt *3d* states. Majority (spin-up) states are fully occupied in both cases, whereas minority (spin-down) states have at least one unoccupied *3d*-band, giving rise to a magnetic moment of at least 1 $\mu_B$. The contribution of Co *4s* electrons to the magnetic moment is negligible, as can be seen from the DOS projected onto *4s* states. The case of the $J_{H,high}$ state is different due to additional on-site Coulomb repulsion applied in the calculations in the form of a Hubbard U-correction. In this case, filling of minority *3d* states is energetically less favorable, leading to a considerable energy separation between the occupied and unoccupied states. This also results in a larger imbalance between the occupied majority and minority states, increasing the magnetic moment of the *3d*-shell. One can also see that cobalt *3d* states are broadened, which may be associated with a stronger hybridization with phosphorus states. It is worth noting that the weight of cobalt states near the valence and conduction band edges for $J_{H,high}$ is substantially smaller than for the $J_T$ and $J_{H,low}$ states. A more detailed energy spectrum calculated for the different cobalt states can be seen in Fig. S7, where *k*-resolved band structure projected onto both majority and minority Co *3d* states is shown.

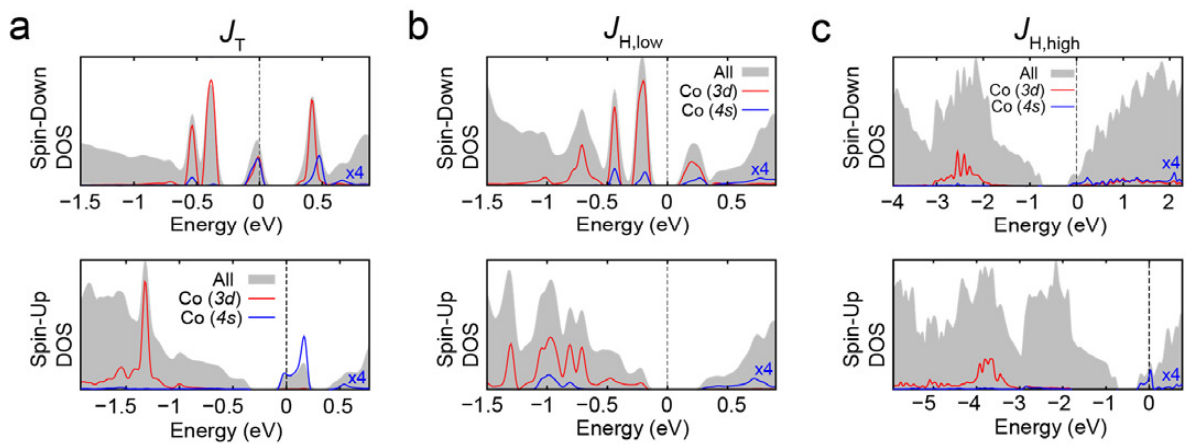

**Supplementary Figure 6. Electronic density of state calculations.** Spin-resolved DOS calculations for Co + BP with $J_T$ (left), $J_{H,low}$ (middle), and $J_{H,high}$ (right). Red curves represent Co *3d* orbital contributions, blue curves show Co *4s* orbital contributions (magnified by a factor of 4), and gray show total DOS.

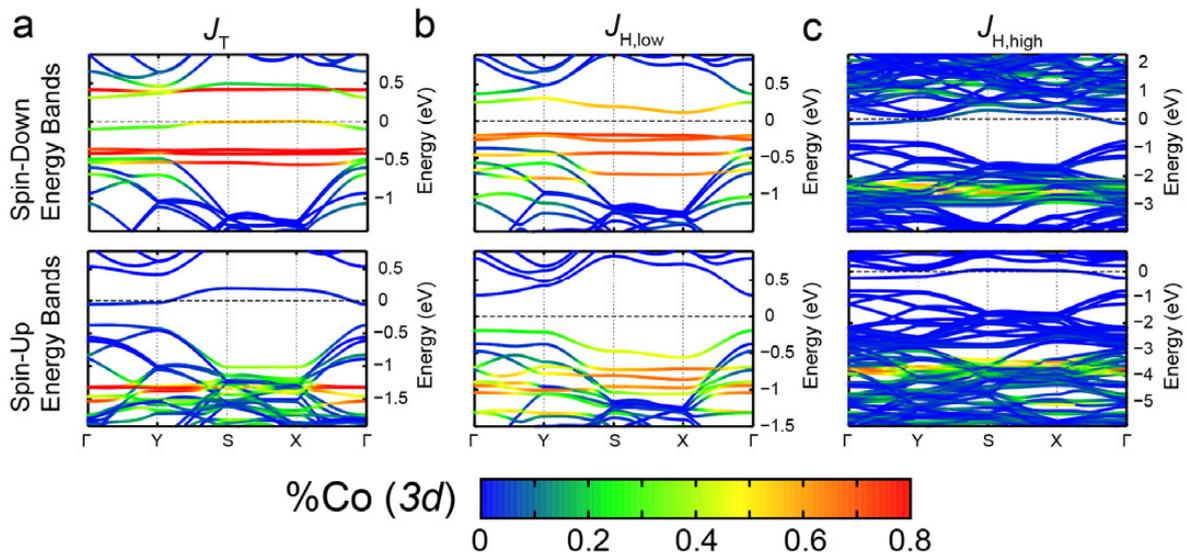

**Supplementary Figure 7. Band structure of Co on BP.** Spin-resolved calculations for Co + BP with $J_T$ (left), $J_{H,low}$ (middle), and $J_{H,high}$ (right). Color scale indicates percentage of Co $3d$-orbital contribution to each band.

In Table S1, we list the spin-resolved $3d$-subshell occupancies of the Co adatom on BP calculated for the two orbital configurations. From the projections, it is clear a major orbital redistribution takes place within the $3d_{xz}$ and $3d_{yz}$ subshells upon vertical relaxation ($\Delta n \approx 0.4$ e$^-$). Additionally, the reduced ligand interactions make the $3d_{z2}$ subshell more favorable as the atom relaxes away from the BP surface. One can see that in the $J_{H,low}$ state the magnetic moment originates predominantly from the $3d_{x2-y2}$ orbital with some contribution from the $3d_{z2}$ orbital. On the contrary, the magnetization in the $J_{H,high}$ state is distributed almost equally over all orbitals of the $3d$ symmetry except $3d_{z2}$, whose contribution is negligible.

|  | $J_{H,low}$ | | | | $J_{H,high}$ | | | |
| --- | --- | --- | --- | --- | --- | --- | --- | --- |
|  | Spin-up $n$ | Spin-down $n$ | Total $n$ | Magnetic Moment | Spin-up $n$ | Spin-down $n$ | Total $n$ | Magnetic Moment |
| $3d_{z^2}$ | 0.945 | 0.785 | 1.73 | 0.16 | 0.991 | 0.962 | 1.953 | 0.029 |
| $3d_{xy}$ | 0.932 | 0.89 | 1.822 | 0.042 | 0.99 | 0.685 | 1.675 | 0.305 |
| $3d_{x^2-y^2}$ | 0.95 | 0.493 | 1.443 | 0.457 | 0.995 | 0.647 | 1.642 | 0.348 |
| $3d_{xz}$ | 0.928 | 0.883 | 1.811 | 0.045 | 0.972 | 0.404 | 1.376 | 0.568 |
| $3d_{yz}$ | 0.944 | 0.873 | 1.817 | 0.071 | 0.986 | 0.496 | 1.482 | 0.49 |
| $3d_{tot}$ | 4.699 | 3.924 | 8.623 | 0.775 | 4.934 | 3.194 | 8.128 | 1.74 |
| $4s$ | 0.124 | 0.117 | 0.241 | 0.007 | 0.266 | 0.116 | 0.382 | 0.15 |
| $4s+3d_{tot}$ | 4.823 | 4.041 | 8.864 | 0.782 | 5.2 | 3.31 | 8.51 | 1.89 |

**Supplementary Table 1. Projections onto the cubic harmonics.** Electronic charge (spin up, spin, down, and total) and magnetic moment projected onto the cubic harmonics for $J_{H,low}$ and $J_{H,high}$. Due to differences in orthogonality between the original *d*-orbital hybrid functions and the cubic harmonics, the total *d*-orbital occupancy is slightly overestimated with respect to the populations given in figure 2.

To gain insight into the charge redistribution around the Co adatom on the BP surface, we calculate the in-plane averaged charge density difference (*Δn*) along the z-direction. Here, *Δn* is defined as *Δn* = $n_{Co+BP}$ − $n_{BP}$ − $n_{Co}$, where $n_{Co+BP}$ is the charge density of BP with the Co adatom, $n_{BP}$ is the charge density of BP without the adatom, and $n_{Co}$ is the charge density of the free Co atom in a S = 3/2 state. The corresponding quantity (*Δn)* is shown in Fig. S8 for the two orbital configurations of the Co atom on BP. One can see that the charge in the adatom tends to be localized toward the surface plane causing perturbations of the charge density within the surface. At the same time, no significant charge transfer is observed between Co and BP. The primary difference between the two states is the spatial extension of the charge redistribution, which is greater for $J_{H,high}$. An area of comparison is highlighted in purple in Fig. S8a, where the *Δn* for the two states is directly compared. These observations are consistent with the subshell projections from table S1 (increased charge in the $3d_{z^2}$ subshell for $J_{H,high}$), and add further evidence (in addition to the broadened linewidths of Fig. S6) for stronger hybridization in $J_{H,high}$.

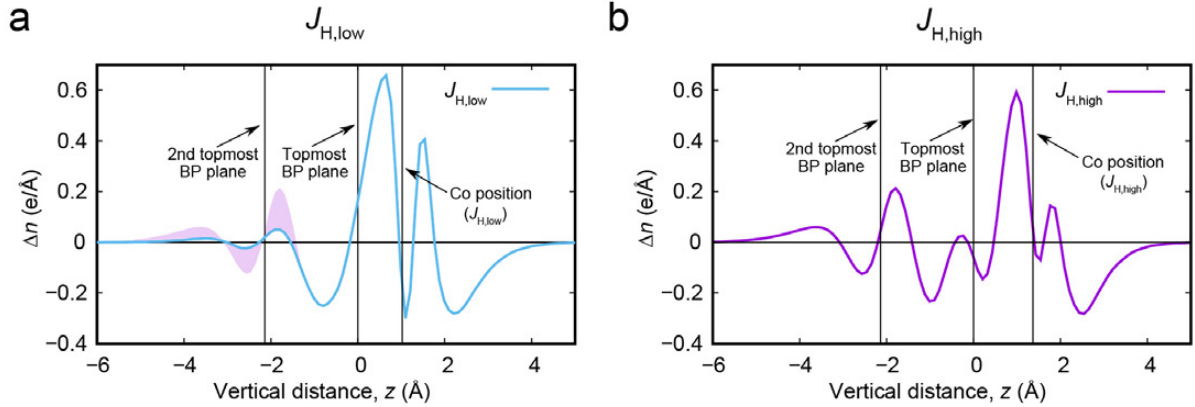

**Supplementary Figure 8. Out-of-plane charge distribution.** Charge redistribution ($\Delta n$) defined in the text for (a) $J_{H,low}$ and (b) $J_{H,high}$. Free Co atom positions are taken from relaxed DFT calculations for $J_{H,low}$ and $J_{H,high}$. Purple shaded area in (a) is a direct comparison of $\Delta n$ for $J_{H,low}$ and $J_{H,high}$ near the lower plane of phosphorus atoms.

Magnetic anisotropy has been calculated for $J_{H,low}$ and $J_{H,high}$ states using DFT taking into account spin-orbit coupling, with the results given in Table 2. The anisotropy is defined as $\Delta E = E_{easy} - E_{hard}$, where $E_{easy}$ and $E_{hard}$ are the total energy calculated when magnetic moments are aligned along the easy and hard axis, respectively. Here $y$ denotes the [100] direction, and $x$ the [010] direction. For both orbital states the hard axis corresponds to the zigzag crystallographic direction [010]. Conversely, the the easy axis for the two states are different, which is the armchair [100] and out-of-plane (z-direction) for $J_{H,low}$ and $J_{H,high}$, respectively. The corresponding anisotropy energies amount to 0.5 meV ($J_{H,low}$) and 0.4 meV ($J_{H,high}$). The fact that the preferential orientation of the magnetic moment is different for the two orbital states is interesting, as this implies the magnetic anistropy can be controlled electrically. The anisotropy energies are generally quite small due to the relatively low-Z phosphorus atoms of the substrate; however, such energies are accessible at temperatures below 1K. Thus spin-averaged and spin-sensitive techniques can potentially be used to investigate the magnetic properties of both $J_{H,low}$ and $J_{H,high}$.

|  | $J_{H,low}$ | $J_{H,high}$ |
|---|---|---|
| *ΔE(z-y)* | 0.31 meV | -0.11 meV |
| *ΔE(x-y)* | 0.49 meV | 0.29 meV |

**Supplementary Table 2. Anisotropy Calculations.** Including spin-orbit coupling into the spin-resolved DFT calculations reveals that the $J_{H,low}$ state has easy plane anisotropy, while the $J_{H,high}$ state has easy axis anisotropy.

**Comparison of d*I*/d*V* spectroscopy with density of states calculations**

Representative d*I*/d*V* spectra for the $J_T$ site (Fig. 2k), $J_{H,low}$ (Fig. 2m), and $J_{H,high}$ (Fig. 2n) all show strong deviation from the bulk BP[5]. Comparing these spectra to calculations for Co on single layer BP, sharp peaks in the d*I*/d*V* spectra can be attributed to Co-related impurity states or ionization events, similar to previous observations in bulk semiconductors[6]. The two peaks observed in the $J_T$ spectrum (Fig. 2k) can be assigned to ionization events, as they shift strongly (>100 mV) with varying tip height (Fig. S9a,d). The primary features in the $J_{H,low}$ spectrum, peaks at -580 mV and 280 mV (Fig. 2m), do not shift significantly with tip height (Fig. S9b,e). The predicted DOS for $J_{H,low}$ shows Co hybrid *d*-bands at both of these energies, which can tentatively be assigned to the origin of the d*I*/d*V* peaks. The measured d*I*/d*V* spectrum for $J_{H,high}$ is given in Fig. 2n. The spectrum is characterized by its diminished intensity and single sharp peak at approximately 420 mV. As seen in Fig. S10g, this peak is related to the ionization of the $J_{H,high}$ state at this bias. The relatively weak total signal and in-gap conduction (Fig. S9c, f) are in good agreement with DFT calculations (Fig. S6c) showing a shift (and reduction) of the band gap with minimal Co DOS in the vicinity of $E_F$.

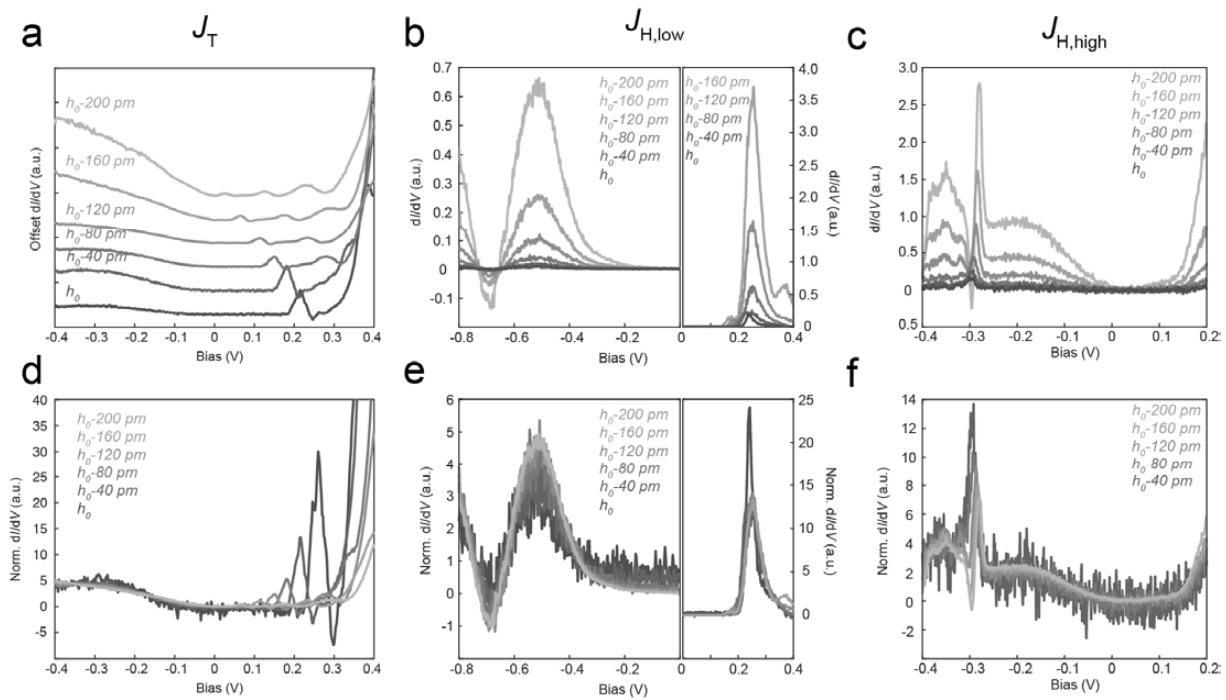

**Supplementary Figure 9. Setpoint-dependent point spectroscopy.** (a) d$I$/d$V$ spectroscopy for the $J_T$ Co atom at various heights indicated in the plot. The nominal height ($h_0$) was set at feedback parameters of $V_s$ = -400 mV, $I_t$ = 20 pA; negative distances are closer to the sample. (b) Two sets of height-resolved d$I$/d$V$ spectroscopy of the $J_{H,low}$ state: one in the valence band, the other in the conduction band. (c) d$I$/d$V$ spectroscopy of the $J_{H,high}$ state at various heights. (d-f) Spectra from (a-c) normalized by a constant proportional to the integrated d$I$/d$V$ signal over the displayed spectral window.

## Tip-induced ionization

Tip-induced band-bending (TIBB) locally influences the energy of semiconductor bands due to insufficient screening from charge carriers; it has been observed that if a defect level, shifted with the material bands, passes through $E_F$, it will undergo an ionization event. The instantaneous change in charge state is observable as a step in the tunneling current (peak in d$I$/d$V$) because the local potential landscape is modified via an effective Coulomb potential at the atom[6-9]. Using d$I$/d$V$ spectral mapping, we map out the spatial location of such a charging event for $J_{H,low}$ and $J_{H,high}$ in Fig. S10a-c, measuring at constant height and the specified bias voltage. The corresponding current images are shown in Fig. S10d-f. The sizes of the isotropic ionization-related events, or charging rings, scale with applied bias (Fig. S10g) roughly following hyperbolic contours of constant TIBB.

As seen in Fig. S10g, $J_{H,low}$ and $J_{H,high}$ show charging rings from 200 mV < $V_s$ < 700 mV, which follow similar scaling as a function of applied bias. From this, we are able to deduce that the local 3$d$-state of the Co atom does not strongly influence the electric-field screening of the tip potential. Fig. S10g also shows the bias dependence of a charging event between -300 mV < $V_s$ < -100 mV observed only for $J_{H,high}$. The event clearly falls on a distinct TIBB contour and as the contour retains positive curvature, the bias range probed (-250 mV to 0 mV) must be above the flat band condition (Fig. 3d) dictated by the tip and sample work functions. Such a condition ($V_{FB}$ ≲ -300 mV) is achievable with a tip work function of 4.0-4.1 eV. Similar flat band conditions were observed with multiple tips after preparation and calibration on Au(111). The interpretation of $V_{FB}$ ≲ -300 mV, is further corroborated by the inverted current response of the $V_s$ < 0 V ionization event (Fig. S10f). As the band bending is still upward in this bias regime, the ionization event is the result of a local state being moved above $E_F$; however, as $V_s$ < 0 V electron current is passing from filled sample states to empty tip states. The loss of one additional filled state (upon ionization) reduces the number of available states contributing to the tunneling current, thereby causing a step-wise decrease in $I_t$ upon ionization. Thus, the ionized region for $J_{H,high}$ at $V_s$ < 0 V appears as a disk-like depression (highlighted by the arrow in Fig. S10f) in the current map, whereas it appears as a disk-like protrusion for both $J_{H,low}$ and $J_{H,high}$ at $V_s$ > 0 V (Fig. S10d,e). Knowledge of the flat band condition necessitates the location of the responsible states below $E_F$, as they must be ionized through upward band bending. This information, in addition to the relative radii for each charging ring, inform the qualitative energy level structure shown in Fig. S10h. From this data, it is clear that Co states near the VB edge (Fig. S6) are ionized through upward TIBB; further quantification of the relevant energies is possible via TIBB simulations[10] extended to account for the BP band structure.

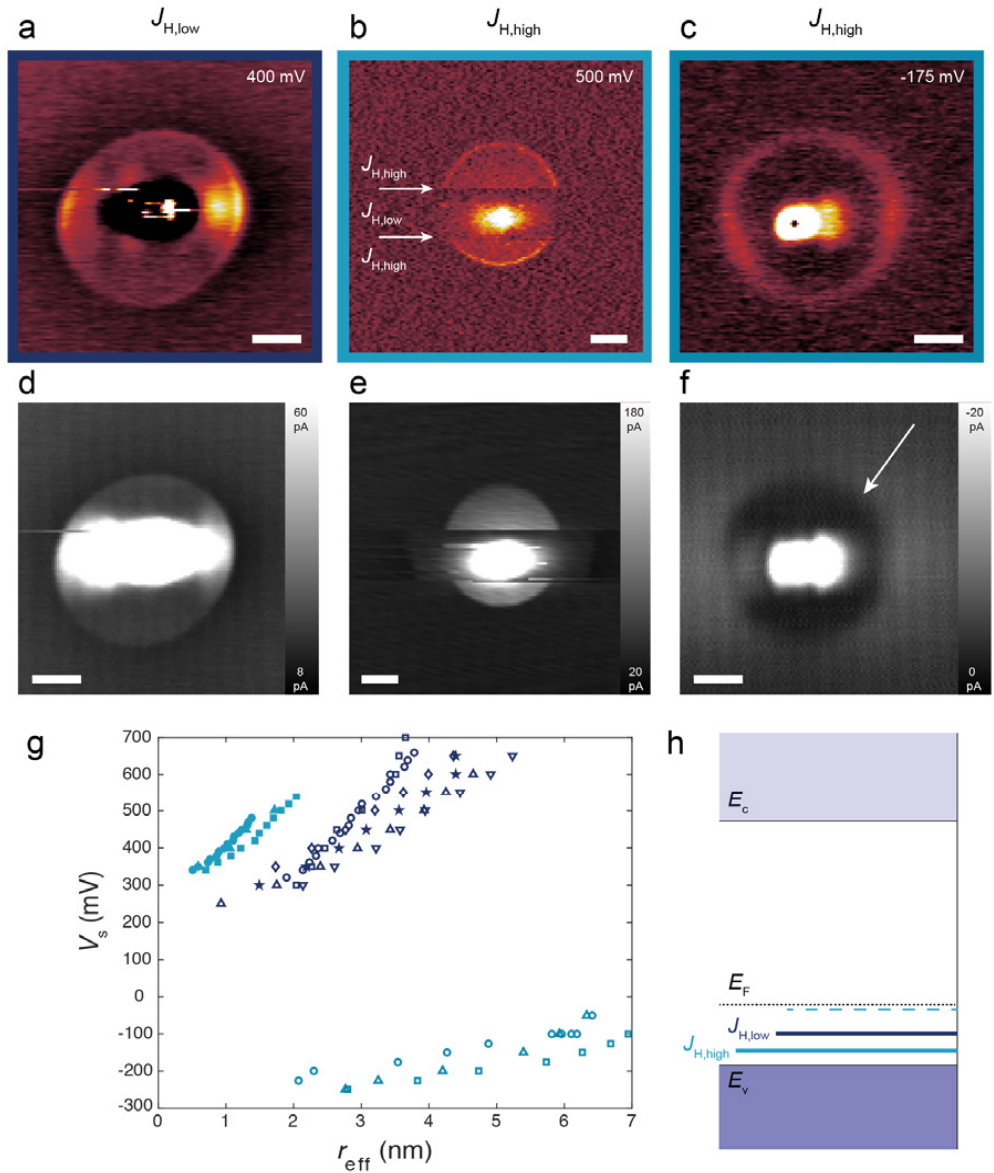

**Supplementary Figure 10. Charging ring maps.** Simultaneously acquired constant-height (a) d$I$/d$V$ and (d) current maps of a $J_{H,low}$ Co atom at $V_s$ = 400 mV. Constant-height (b) d$I$/d$V$ and (e) current maps of the $J_{H,high}$ state at $V_s$ = 500 mV. Constant-height (c) d$I$/d$V$ and (f) current maps of the $J_{H,high}$ state at $V_s$ = -175 mV. All images: scale bar = 1 nm. The decrease in current magnitude in (f), highlighted by the white arrow, is qualitatively understood as the result of a de-charging event at negative sample bias (Fig. 4e); at $V_s$ < 0 V, electrons tunnel from filled sample states into empty tip states, when a filled electronic state is depopulated, the tunneling current decreases. Such an explanation successfully accounts for the isotropic character of the charging event and the spatial evolution of the charging state with decreasing bias magnitude. (g) Evolution of charging ring radii ($r_{eff}$) with bias voltage for $J_{H,low}$ (dark blue) and $J_{H,high}$ (light blue). (h) Qualitative level structure in the flat band condition showing primary $J_{H,low}$ (solid dark blue line) and $J_{H,high}$ (solid light blue line) energy levels responsible for ionization in the regime 200 mV < $V_s$ < 700 mV. Significantly, these are the two states that are ionized during the telegraph switching. An additional $J_{H,high}$ state (dashed light blue line) is depicted, which is responsible for the charging event at -300 mV < $V_s$ < -100 mV.

Figure S11 demonstrates the methods used to calculate the uncertainty when estimating the charging ring radii shown in Fig. 3c. First, the ring radius was approximated via $r_{eff}=L/2\pi$, where $L$ is defined at the circumference of the ring. This effective definition was used to accommodate deviations from ideal circular behavior. The error bars were then determined by one of two methods, shown in Fig. S11. The first method fit the largest and smallest possible ellipsoids to the data, calculating $r_{min}$ and $r_{max}$ as with $r_{eff}$. The second method directly measured the smallest and largest radius for each ring. The larger of the two error values was then displayed using error bars in Fig. 3c.

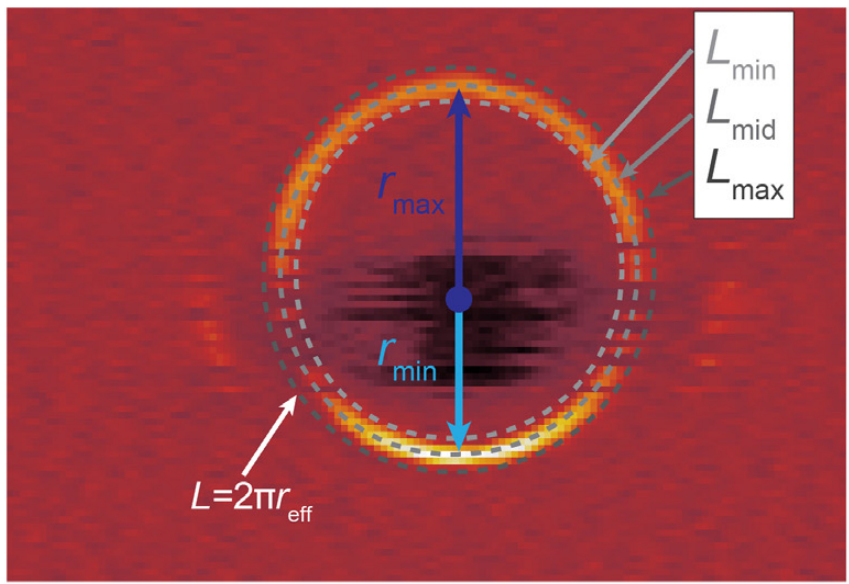

**Supplementary Figure 11. Charging ring uncertainty.** Image of charging ring observed in d$I$/d$V$ signal with two sources of experimental uncertainty shown. One source of uncertainty comes from determining the ring circumference. To estimate the possible error, the minimum and maximum possible circumferences were determined, labeled $L_{min}$ and $L_{max}$ in the figure, and used to correspondingly calculate $r_{min}$ and $r_{max}$ for a given charging ring. The second source of experimental uncertainty was introduced due to a non-spherical tip shape corresponding to a non-circular charging ring. Here, the $r_{min}$ and $r_{max}$ values were directly measured. The larger of the values from the two methods was used for error estimates.

**Room temperature annealing**

First, a sample was dosed with Co and measured at 4.4 K (Fig. S12a – areal density of Co atoms 0.097 nm$^{-2}$). Annealing was then accomplished by removing the sample from the low-temperature STM and inserting it into a UHV sample carousel kept at room temperature for precisely 1 minute. Afterward, the sample was returned into the STM. Subsequent characterization (Fig. S12b – areal density of Co atoms 0.087 nm$^{-2}$) was conducted at 4.4 K. With the exception of some (>40%) hydrogenation (highlighted with a red circle in Fig. S12b), a large portion of the atomic species remain in either the $J_T$, $J_{H,low}$, or $J_{H,high}$ states after the anneal. A second, subsequent anneal was performed by again placing the sample into the room temperature carousel, however, the duration was increased to 10 minutes. After the second annealing, the areal density of surface species was reduced to 0.035 nm$^{-2}$ (Fig. S12c). This reduction in atomic density in conjunction with an increase in the apparent height of many species leads to the conclusion that a sizable portion of the Co atoms formed clusters. However, a number of Co species remained isolated after the annealing procedure (Fig. S12b); two $J_T$ atoms are highlighted with orange boxes (in total, at least six can be found in the image) and two $J_{H,high}$ atoms are indicated with dark red boxes (in total, at least 13 are found in the image). Expanded images showing each of these selected Co species are given at the bottom of Fig. S12c. The formation of Co clusters in Fig. S12c is partially attributed to the high Co density before the annealing step.

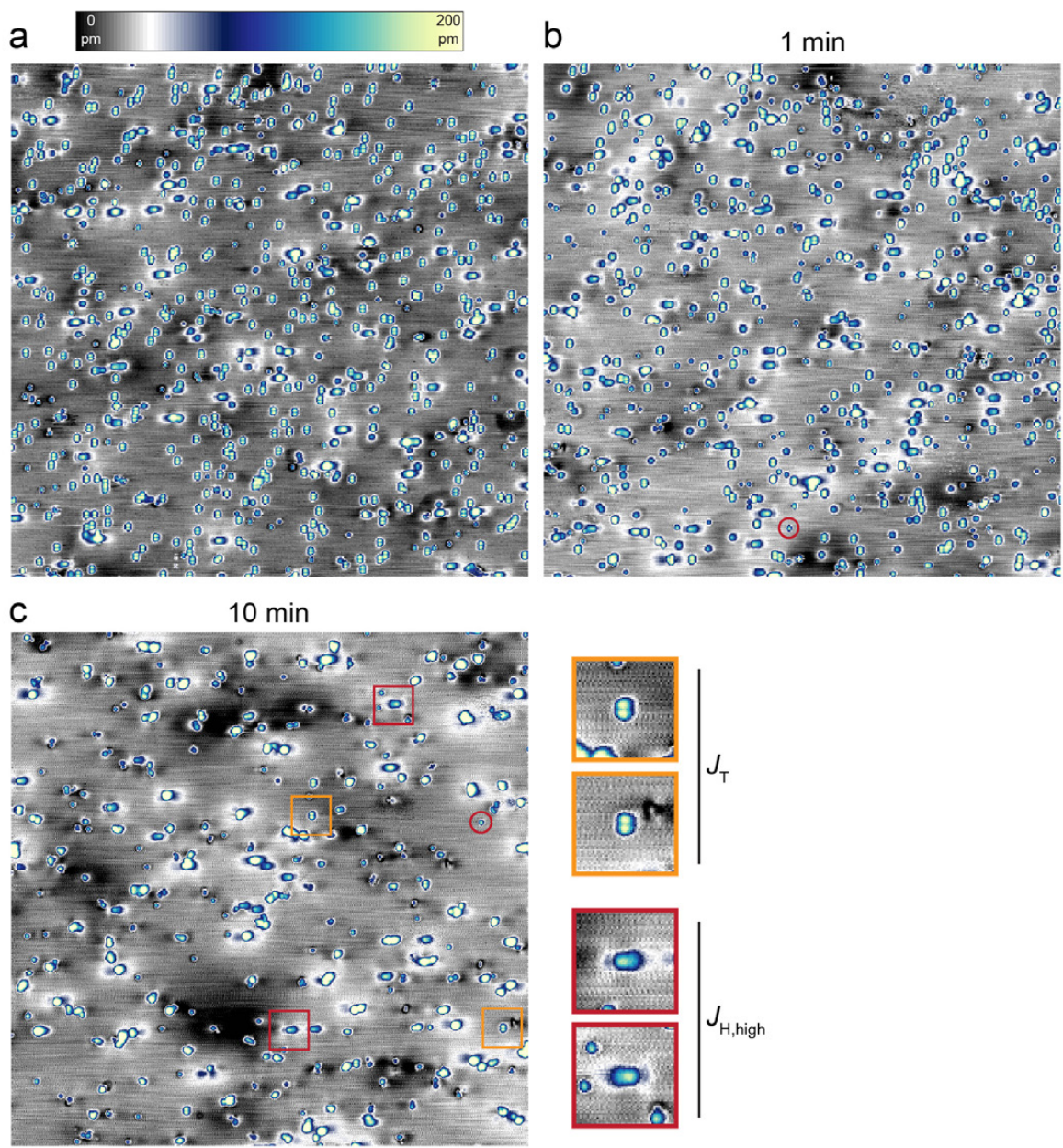

**Supplementary Figure 12. Room temperature annealing.** (a) Large area image of higher coverage (0.097 nm$^{-2}$) Co species acquired at 4.4 K before annealing ($V_s$ = -400 mV, $I_t$ = 20 pA, image = 80 nm x 80 nm). (b) Sample after annealing for 1 minute and cooling back to 4.4 K ($V_s$ = -400 mV, $I_t$ = 20 pA, image = 80 nm x 80 nm). (c) Sample after annealing for 10 minutes and cooling back to 4.4 K ($V_s$ = -400 mV, $I_t$ = 20 pA, image = 80 nm x 80 nm). Enlarged images at the bottom of (c) show isolated $J_T$ (orange boxes) and $J_{H,high}$ (red boxes) species that remain after room temperature annealing.